# Evaluation of post-blast damage in cut blasting with varying extra-depths: insights from 2D simulations and 3D experiments

*Changda Zheng[1], Renshu Yang[1,2], Jinjing Zuo[2], Canshu Yang[1], Yuanyuan You[3], and Zhidong Guo[4]*

1) School of Future Cities, University of Science and Technology Beijing, Beijing 100083, China

2) Research Institute of Macro-Safety Science, University of Science and Technology Beijing, Beijing 100083, China

3) School of Mechanics and Civil Engineering, China University of Mining and Technology (Beijing), Beijing 100083, China

4) Ansteel Cornerstone Mining Corporation, Limited, Anshan 114047, China

**Abstract:** In blasting engineering, borehole utilization is a key metric for evaluating blasting performance. While previous studies have examined the effects of expansion space, cutting design, in-situ stress conditions, and rock properties on borehole utilization, research on the intrinsic relationship between extra-depth—defined as the portion of the cut hole extending beyond the depth of auxiliary holes—and borehole utilization remains insufficient. This gap in understanding has hindered the resolution of issues such as residual boreholes and unbroken rock at the borehole bottom in deep-hole blasting, thereby limiting improvements in borehole utilization. This study employs a simplified double-hole model for extra-depth cut blasting to conduct two-dimensional numerical simulations and three-dimensional cement mortar model experiments. It systematically investigates the blasting damage characteristics, fractal damage, and energy evolution under varying extra-depth as a single variable. Experimental parameters such as borehole utilization, cavity diameter, cavity volume, and fragment size distribution were obtained to comprehensively analyze the nonlinear effects of extra-depth on post-blast rock damage and its mechanisms. Both simulation and experimental results indicate that blasting damage parameters exhibit a nonlinear trend of initially increasing and then decreasing with increasing extra-depth. Appropriately increasing the extra-depth improves rock breakage efficiency, while excessive extra-depth reduces efficiency due to confinement effects at the borehole bottom. Adjusting the extra-depth can optimize the distribution of explosive energy between rock fragmentation and rock ejection. Additionally, numerical simulations reveal that the peak internal energy closely aligns with the fractal damage evolution, providing a novel perspective for quantitatively assessing post-blast damage. The findings offer scientific guidance for optimizing extra-depth parameters in blasting design and propose a new method for the rapid assessment of blasting damage, enhancing both efficiency and safety in resource extraction operations.

---

[1]Corresponding author: Changda Zheng    E-mail address: zhengcd95@163.com
School of Future Cities, University of Science and Technology Beijing 2024

# 1 Introduction

Cut blasting, a commonly used blasting method for the removal or destruction of hard materials such as rock and concrete in mining, tunneling, and construction projects, is influenced by various factors. Researchers and engineers need to comprehensively consider and optimize these factors to enhance blasting effectiveness and safety. The factors that affect the effectiveness of cut blasting include, but are not limited to: (1) free surface [1, 2] and expansion space [3, 4], which facilitate the transfer of blasting energy, increase the degree of rock fragmentation, and significantly improve blasting results; (2) the physical and mechanical properties of the rock [5], including hardness, strength, toughness, structure, and bedding[6], all of which directly impact the blasting effect; (3) tectonic stress [7, 8], as the magnitude and direction of in-situ stresses play a crucial role in determining blasting outcomes; (4) the type and quantity of explosives [9, 10], as different explosives exhibit distinct variations in explosive energy, detonation speed, and gas release, with the amount of explosives directly influencing blasting power, both of which significantly affect the blasting result; (5) cutting design parameters such as borehole length [11], diameter [12], spacing, angle, charge structure [13, 14], initiation position, delay time [15, 16], and cutting configuration [17, 18], all of which have a noticeable impact on the blasting performance.

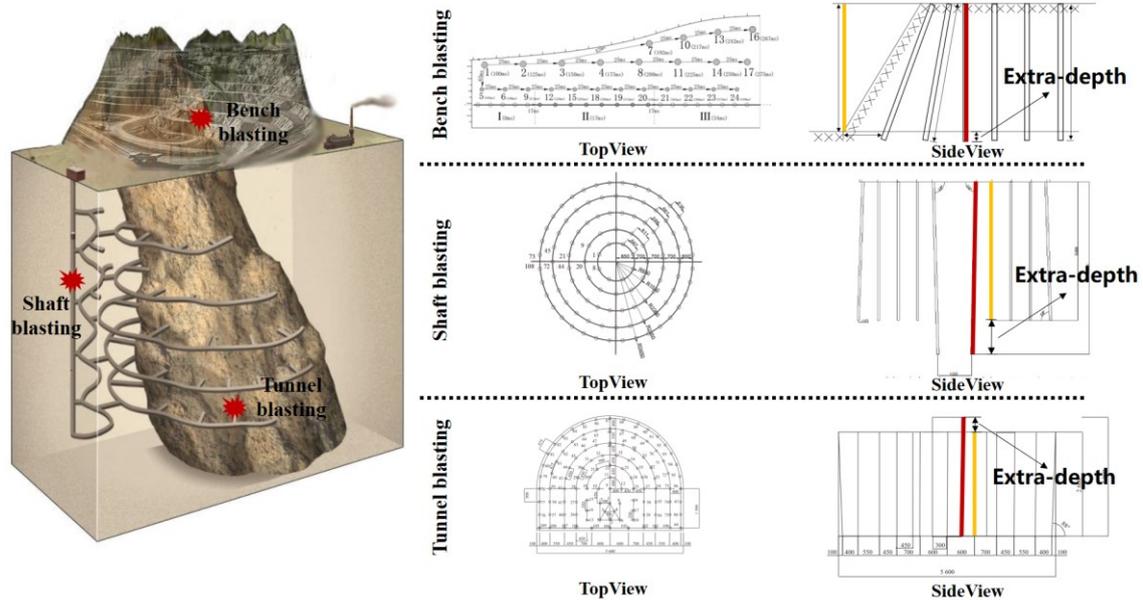

Fig.1 Extra-depth in bench blasting, shaft blasting, and tunnel blasting

Among these factors, extra-depth—defined as the portion of the cut hole extending beyond the depth of auxiliary holes—is also a crucial parameter affecting blasting effectiveness. Its primary role is to overcome the confinement effect in cut blasting, thereby improving the overall blasting performance. Fig.1 illustrates the concept of extra-depth in bench blasting, shaft blasting, and tunnel blasting. In shaft and tunnel blasting, extra-depth refers to the portion of the cut hole that exceeds the depth of the auxiliary hole. According to the book blasting design and construction, in tunnel excavation blasting, the cut hole should be 150mm to 250mm deeper than the auxiliary holes; in shaft excavation blasting, the cut hole should be 200mm to 300mm deeper than the auxiliary holes. In bench blasting, extra-depth refers to the portion of the drill hole that exceeds the elevation of the bench bottom. The purpose is to lower the center of the charge, overcoming the confinement effect of the rock at the bottom of the bench, thus minimizing or eliminating the formation of a bottom ledge after blasting and ensuring a flat platform. For soft rocks, a smaller value is selected, while for hard rocks, a larger value is used. In domestic mines, the typical extra-depth value ranges from 0.25m to 3.6m. It is important to note that the existing recommendations for extra-depth are largely based on shallow-hole blasting experiences, which are increasingly inadequate to meet the demands of deep-hole blasting technology. In recent years, with the increasing depth of mineral resource extraction and the advancement of mechanization, deep-hole blasting techniques have become more widely applied. In the 1970s, the average depth of single-cycle boreholes in Chinese shaft excavations was 2.93m, but today it has increased to 4.31m, with some shafts reaching depths of 5m to 6m. Similarly, in the 1990s, the majority of single-cycle borehole depths in coal mine tunnels were below 2.0m, accounting for 56.25%, but now, blasting schemes with single-cycle borehole depths under 2.0m only account for 34.69%, with some coal mine tunnels reaching depths above 3.0m. Given these developments, it is necessary to reconsider the rationality of determining extra-depth based on the design practices of shallow-hole blasting.

Currently, research on extra-depth blasting is largely empirical and lacks systematic and quantitative theoretical guidance. Excessive extra-depth not only leads to waste of drilling and explosives but also increases the damage to the subsequent excavation face, making it more difficult to drill the next hole. On the other hand, insufficient extra-depth may leave residual holes at the excavation face, affecting blasting quality and work efficiency. This highlights the complexity and necessity of optimizing the extra-depth parameter for cut blasting. Given this, how to scientifically design a reasonable extra-depth remains a critical issue. Especially under deep-hole blasting conditions, blasting designs lacking scientific guidance may lead to economic losses and safety risks. There is an urgent need to reveal the intrinsic mechanism of extra-depth's impact on post-blasting rock damage. Therefore, this paper aims to systematically explore the mechanisms by which extra-depth influences blasting damage and propose a quantifiable method for rapid damage assessment.

To address these issues, this study conducts two-dimensional numerical simulations and three-dimensional cement-sand model experiments using a simplified dual-hole model for extra-depth cut blasting. The research systematically investigates the blasting damage characteristics, fractal damage levels, and energy evolution patterns as the extra-depth varies as a single parameter. The experiments provide data on post-blasting parameters such as hole utilization, cavity diameter, cavity volume, and fragment size distribution, offering a comprehensive analysis of the nonlinear effects of extra-depth on post-blasting rock damage and its underlying mechanisms. The results not only provide a theoretical basis for optimizing extra-depth parameters in cut blasting but also offer a new method for rapid blasting damage assessment. This work is of significant theoretical and economic value for improving mining resource development efficiency and safety.

## 2 Two-dimensional numerical simulation of cut blasting with different extra-depth

### 2.1 Establishment of numerical model and design of case

Chenxi Ding et al. [19] demonstrated that in three-dimensional numerical simulations, the fractal damage curves of two mutually perpendicular two-dimensional cross-sections exhibit consistent variation trends. Although two-dimensional numerical simulations have certain limitations, they still provide valuable insights into the mechanisms of rock damage caused by explosive loading. Therefore, this section simplifies the inherently three-dimensional problem of

Using LS-DYNA, five numerical models of cut blasting with varying extra-depth were designed, as shown in Fig.2. These models, labeled Case-1, Case-2, Case-3, Case-4, and Case-5, use $H_e$——the cut hole beyond the depth of the non-cut holes as the sole variable across the cases. The simulation results were analyzed and compared from two perspectives: post-blast damage distribution and energy evolution during the blasting process. The extra-depth for Case-1 to Case-5 were set at 0 mm, 5 mm, 10 mm, 15 mm, and 20 mm, respectively. The model dimensions were 400 mm × 400 mm, with a thickness of one mesh element. In all five models, the charge length of the cut hole was set to 35 mm, while the stemming length of the non-cut holes was 15 mm, with a charge length of 35 mm. The initiation point was located at the geometric center of the bottom of the explosive, and the detonation sequence was designed such that the I segment of the cut hole detonated first, followed by the II segment of the non-cut holes, with a delay interval of 25 ms.

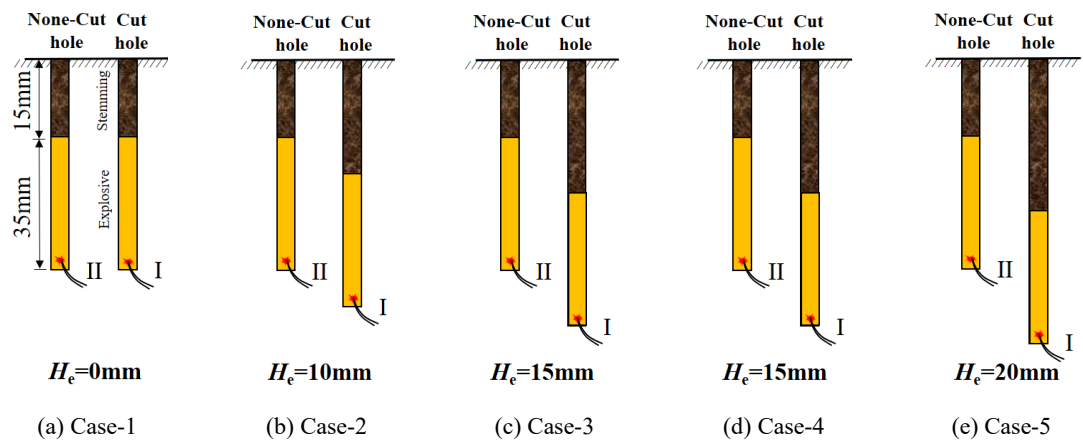

Fig.2 Numerical simulation design for Case-1~Case-5

The explosive material was defined using the keyword *MAT_HIGH_EXPLOSIVE_BURN, and the rock mass constitutive model was defined using the keyword *MAT_RHT. The parameters for both the explosive and the rock mass were set based on the reference [20]. The Jones-Wilkins-Lee (JWL) equation of state for the explosive is as Equation (1), describing the relationship between pressure and specific volume during the detonation process：

$$P = A\left(1 - \frac{\omega}{R_1 V}\right)e^{-R_1 V} + B\left(1 - \frac{\omega}{R_2 V}\right)e^{-R_2 V} + \frac{\omega E_0}{V} \quad (1)$$

where: $A$, $B$, $R_1$, $R_2$, and $\omega$ are material constants; $P$ represents the pressure; $V$ denotes the relative volume of detonation products; and $E_0$ is the initial specific internal energy of the detonation products.

The boundary conditions were configured to reflect those in rock mass cut blasting scenarios. The top surface of the model was left unconstrained to simulate a single free surface during cut blasting. The left, right, and bottom surfaces were set as non-reflecting boundaries, while the front and rear surfaces were assigned fixed constraints. A fluid-solid coupling approach was employed to establish the interaction between the explosive and the rock mass. In this configuration, the explosive and air regions were defined as fluids, whereas the rock mass was defined as a solid.

## 2.2 Analysis of blasting damage evolution

The numerical simulation results of post-blast damage for Case-1 to Case-5 are presented in Fig.3. As observed in Fig.3, the post-blast damage in Case-1 exhibits a certain degree of symmetry. However, with the downward shift of the explosive charge position in the cut holes, the blast-induced damage in Case-2 to Case-5 displays asymmetry, which becomes increasingly pronounced with the increase in extra-depth. These observations are limited to qualitative descriptions. As for the quantitative characterization of the extent of damage, it cannot be visually assessed directly from the images.

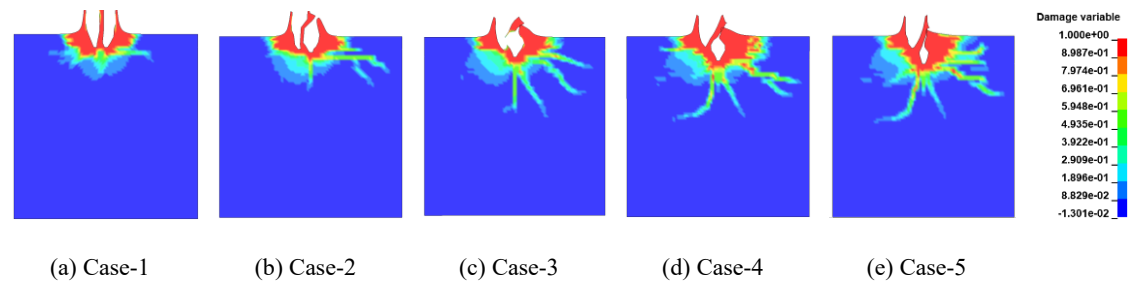

(a) Case-1　　(b) Case-2　　(c) Case-3　　(d) Case-4　　(e) Case-5

Fig.3 Numerical simulation results of post-blast damage for Case-1~Case-5

To further quantitatively evaluate blast-induced damage, an equivalent method for simulating fractures in rock based on finite element numerical results, as proposed by Changping Yi et al. [20], was adopted. In this approach, elements with a damage level exceeding 0.7 are removed, thereby approximating the blast damage derived from finite element simulations. Using this method, 0 and 0.7 were set as the minimum and maximum values of the display range, respectively, to visualize the post-blast damage for Case-1 to Case-5. The results of the numerical simulations are illustrated in Fig.4.

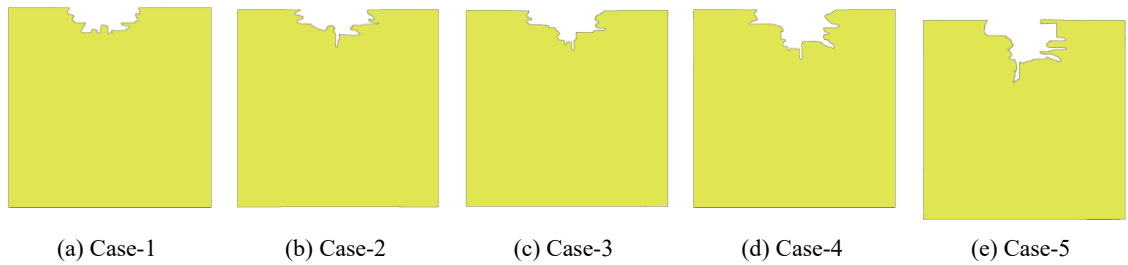

| (a) Case-1 | (b) Case-2 | (c) Case-3 | (d) Case-4 | (e) Case-5 |

Fig.4 Post-blast damage of numerical simulations from Case-1~Case-5

The dual complexities of rock constitutive behavior and explosive dynamics significantly increase the difficulty of quantitatively evaluating blast-induced damage in rock masses. Consequently, developing objective measures of blast damage and establishing a reliable blast damage evaluation system remain critical challenges in the field of blasting engineering. Researchers have proposed various damage variables to characterize blast-induced damage in rock masses, including fractal dimensions [21-24], micro-cracks density [25], variations in equivalent elastic modulus [26], and changes in the medium's P-wave velocity [27]. Among these, the application of fractal theory has gained considerable attention due to its ability to quantify damage under explosive loading. This approach is particularly appealing as the distribution characteristics of blast-induced cracks, fragment size distributions, and the roughness of fracture surfaces all exhibit fractal properties under explosive loading.

This study employs fractal theory to quantitatively evaluate post-blast damage in numerical simulations. Following the fractal evaluation method for blast damage outlined in reference [2], MATLAB was used to binarized the post-blast damage results of Case-1 to Case-5 (Fig.4), yielding binary images of post-blast damage for Case-1 to Case-5 as shown in Figure 5. Subsequently, MATLAB was utilized to calculate fractal dimensions and conduct fractal damage analysis on the binary images. The results of the fractal dimension calculations and fractal damage analyses are presented in Fig.6 and Fig.7, respectively. As shown, the fractal damage for Case-1 to Case-5 exhibits a trend of initially increasing and then decreasing with the increase in extra-depth. The peak fractal damage in the numerical simulations is observed in Case-4, corresponding to an extra-depth of 15 mm, with a fractal damage value of 0.853.

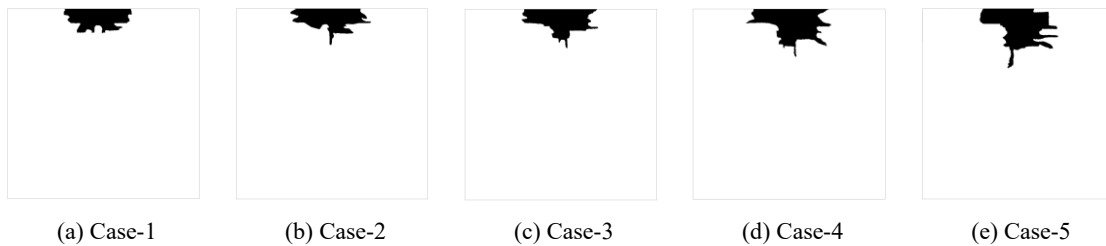

| (a) Case-1 | (b) Case-2 | (c) Case-3 | (d) Case-4 | (e) Case-5 |

Fig.5 Binary images of post-blast damage from numerical simulations in Case-1~Case-5

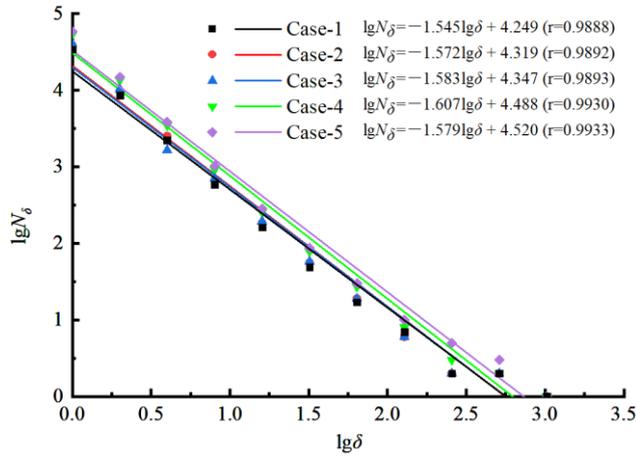

Fig.6 Fractal dimension fitting lines of post-blast damage in Case-1~Case-5

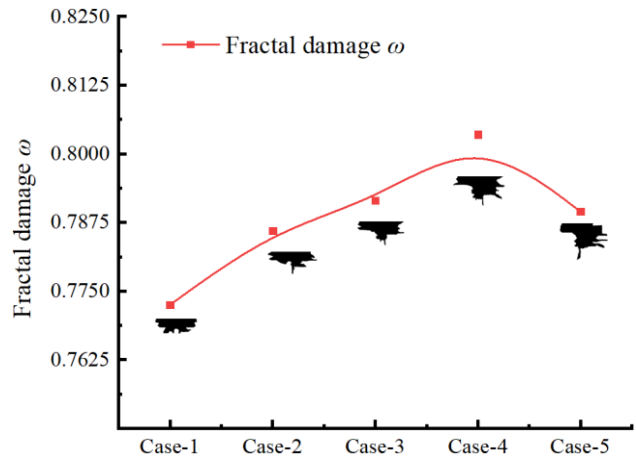

Fig.7 The fractal damage ω variation curve with Case-1~Case-5

## 2.3 Analysis of energy evolution during the blasting process

The three types of energy data—kinetic energy, internal energy, and total energy—of the rock portion during the blasting process were extracted. In the numerical simulation, part of the energy generated by the explosive is transferred to the rock. The energy transferred includes one portion that imparts velocity to the rock, referred to as kinetic energy, and another portion that either causes the rock to fracture or dissipates within the rock, referred to as internal energy. The sum of kinetic energy and internal energy equals the total energy. Since the explosive parameters and geometric dimensions in Case-1 to Case-5 are entirely identical, it can be assumed that the total energy generated by the explosive is the same for all five cases.

Fig.8 shows the variation curves of total energy over time for Case-1 to Case-5. Total energy represents the overall amount of energy consumed by the explosive during blasting to perform work on the rock. As shown in the figure, at the later stage of the explosion, specifically at $t=500$ μs, the total energy of the five cases ranks as follows: Case-5 > Case-4 > Case-3 > Case-2 > Case-1. With the increase in extra-depth, the amount of energy transferred from the explosive to the rock increases correspondingly. The underlying reason for this phenomenon can be explained as follows. Assuming a charge is detonated within a rock mass at infinite depth, the rock, constrained by the infinite depth, cannot move. Under such conditions, the energy generated by the explosion is entirely transferred into the rock as internal energy. At this point, the total energy in the rock reaches its theoretical maximum, which equals the total energy generated by the explosive, entirely dissipated in the form of internal energy. In essence, the energy generated by the explosive is entirely absorbed by the rock. Thus, as the extra-depth increases and the charging position moves downward, the total energy within the rock naturally increases.

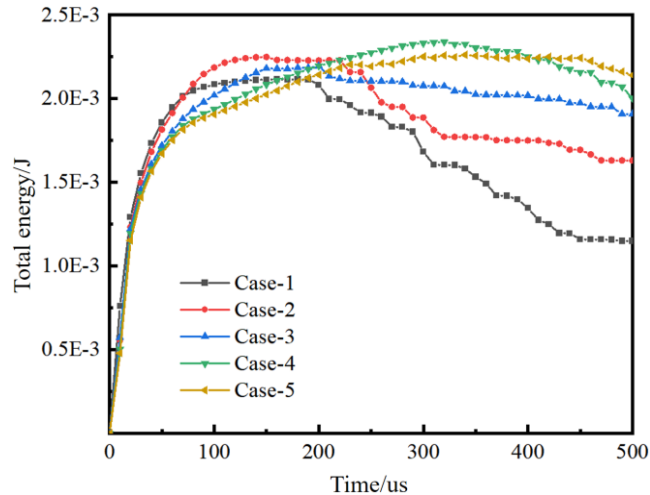

Fig.8 Relationship curves between total energy and time for Case-1~Case-5

Fig.9 illustrates the variation curves of kinetic energy over time for Case-1 to Case-5. As observed from the figure, when the explosion time is less than approximately 270 μs, the extra-depth is inversely proportional to the kinetic energy in the rock. This phenomenon can be explained as follows: when the extra-depth is small, the distance for the explosive stress wave to reach the free surface is shorter, resulting in less attenuation. Consequently, the energy contained in the stress wave remains high, allowing it to cause significant tensile damage through reflection and eject a large quantity of surface rock. Therefore, when the extra-depth is small, the kinetic energy in the rock is higher. Conversely, when the depth of the cut hole is greater, the stress wave travels a longer distance to reach the free surface, leading to greater attenuation and lower energy in the stress wave, making it difficult or even impossible to damage the surface rock through reflection. Hence, when the extra-depth is large, the kinetic energy in the rock is lower. When the explosion time exceeds 270 μs, the kinetic energy in the rock behaves differently: Case-1, with the smallest extra-depth, exhibits the lowest kinetic energy, while Case-5, with the largest extra-depth, has the highest kinetic energy. The reasoning is as follows: when the extra-depth is smallest, the rock mass above the charge is lighter, and the time required for the explosive stress wave to reach the free surface is shorter. As a result, the kinetic energy in the rock initially increases rapidly but also decays quickly. After 270 μs, the kinetic energy in Case-1 has significantly diminished, entering a deceleration phase. In contrast, the kinetic energy in Case-5 is still in the growth phase, as the greater extra-depth delays the full development of kinetic energy in the rock.

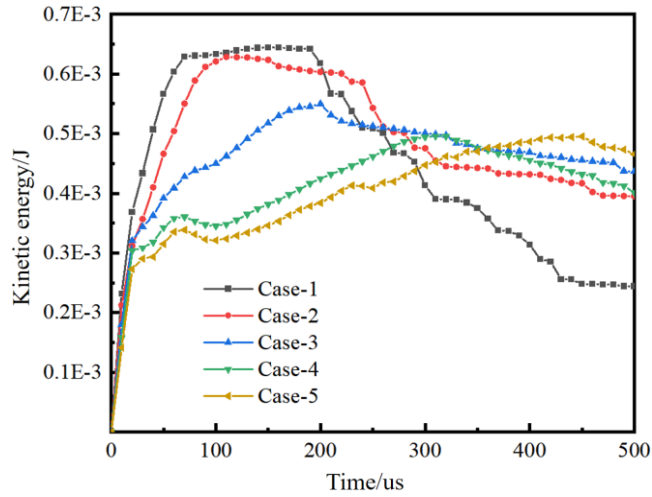

Fig.9 Relationship curves between kinetic energy and time for Case-1~Case-5

Fig.10 presents the variation curves of internal energy over time for Case-1 to Case-5. Internal energy represents the portion of the total energy that causes rock fracture or dissipates within the rock, meaning it is the energy absorbed by the rock from the explosive. This energy manifests in two forms: one is macroscopic damage, visible as cracks, and the other is microscopic damage, imperceptible to the human eye, represented by particle vibration. Therefore, in LS-DYNA, internal energy can be selected as an indicator to evaluate blasting damage, where higher internal energy corresponds to greater damage. As shown in the figure, in the later stages of blasting ($t=500$ μs), the internal energy follows the order Case-4 ≈ Case-5 > Case-3 > Case-2 > Case-1. With increasing extra-depth, the blasting damage to the rock initially increases and then gradually stabilizes. Combining these results with those of the model experiments, it can be concluded that appropriately increasing the extra-depth enhances the damage to the blasted rock mass without increasing the explosive amount. This effectively improves the overall borehole utilization and maximizes the rock-breaking capacity of the cut hole.

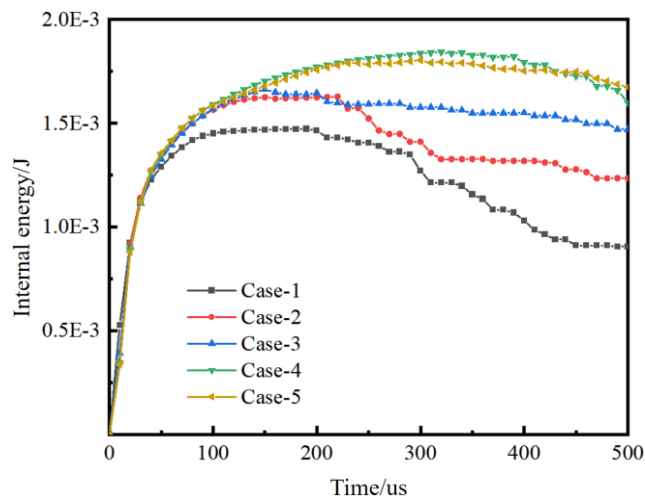

Fig.10 Relationship curves between internal energy and time for Case-1~Case-5

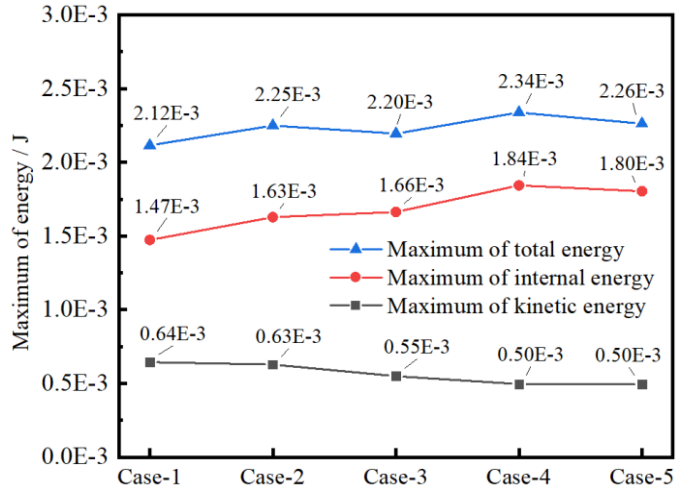

Fig.11 Maximum of total energy, internal energy, and kinetic energy in Case-1~Case-5

Fig.11 illustrates the variation trends of the maximum values of total energy, internal energy, and kinetic energy in Case-1 to Case-5. It can be observed that as the extra-depth increases, the maximum total energy ranges from $2.12×10^{-3}$ J to $2.34×10^{-3}$ J, showing an initial increase followed by a decrease. Notably, Case-4 reaches the peak total energy of $2.34×10^{-3}$ J. Similarly, the maximum internal energy exhibits a comparable trend, ranging from $1.47×10^{-3}$ J to $1.84×10^{-3}$ J, with a peak also occurring in Case-4. In contrast, the maximum kinetic energy is significantly lower than both the total and internal energies, ranging from $0.50×10^{-3}$ J to $0.64×10^{-3}$ J. As the extra-depth increases, the kinetic energy gradually decreases and stabilizes in Case-4 and Case-5.

Furthermore, Fig.11 validates the feasibility of using the maximum internal energy to assess the degree of blast-induced damage. In general, rock blasting damage refers to the phenomenon in which the shock waves and high-pressure gases generated by the explosion of explosives exert loads on the rock, causing the initiation, propagation, and coalescence of micro-cracks within the rock, ultimately forming macroscopic fractures or fragmented zones. Comparing the definition of internal energy in this study with the understanding of rock blasting damage described in prior literature, it is evident that the two concepts align closely. Additionally, by comparing the trends in the fractal damage curves and the maximum internal energy curves for Case-1 to Case-5, a consistent pattern is observed between the two. This suggests that, in addition to using fractal damage to evaluate post-blast damage, the maximum internal energy in numerical simulations can serve as an approximate indicator of the degree of blast-induced damage. This finding provides a novel perspective for evaluating blast damage in numerical simulations.

## 3 Three-dimensional model experiments on cut blasting with different extra-depth

### 3.1 Experimental design

Fig.12 shows the experimental flowchart for the extra-depth cut blasting tests using cement mortar models. Cement mortar was prepared with a mix ratio of cement: sand: water = 1:7.25:1.55, and 10 standard mortar specimens measuring 50 mm × 100 mm, along with 5 models which were labeled as M-1, M-2, M-3, M-4, and M-5, were cast. The basic physical and mechanical parameters of the standard mortar specimens, including density, uniaxial compressive strength, tensile strength, and P-wave velocity, were then measured. Simultaneously, blasting experiments on the models were conducted according to the designed plan. After the experiments, the post-blast cavity depth, cavity diameter, and cavity volume of the models were measured, and the fragmented blocks were sieved for analysis.

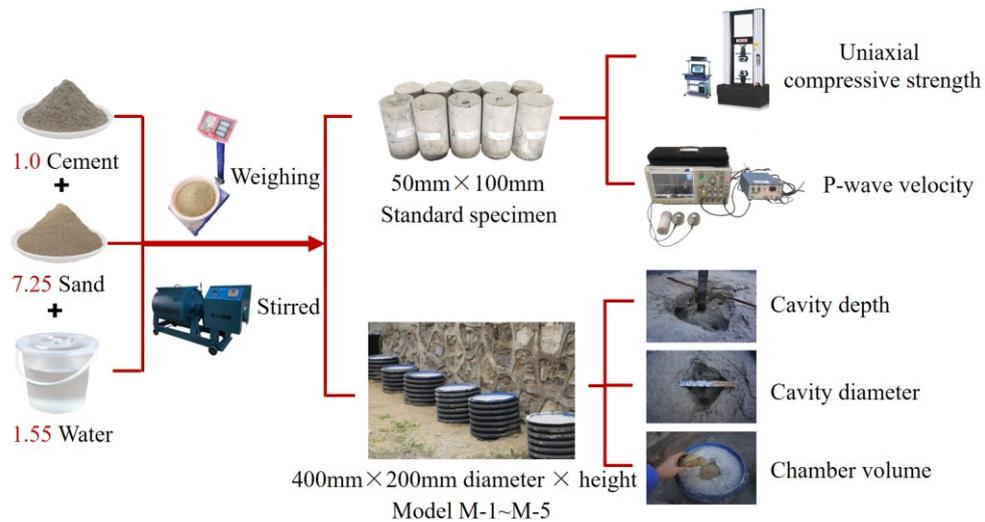

Fig.12 Flowchart of three-dimensional model experiment

Fig.13 illustrates the experimental design for the extra-depth cut blasting tests using cement mortar models. The extra-depth——$H_e$ was set as the sole variable, with extra-depths for models M-1, M-2, M-3, M-4, and M-5 designed as 0 mm, 5 mm, 10 mm, 15 mm, and 20 mm, respectively. Other parameters were identical across the five models: the depth of the non-cut holes was uniformly 50 mm, the explosive structure was a continuous charge, the charge length was 35 mm, and the explosive charge was 1.70 g per hole. The initiation position was set as bottom initiation, with the initiation sequence set as the I cut hole detonating first, followed by the II non-cut hole. The delay interval was set as 25 ms.

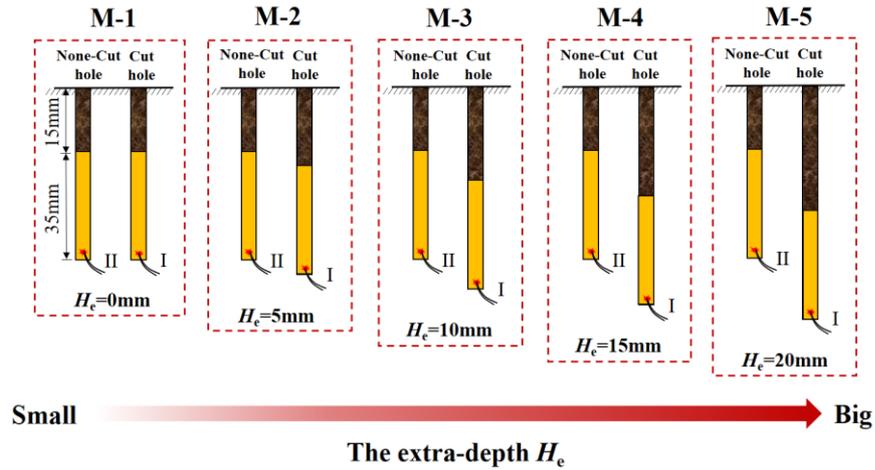

Fig.13 Scheme of three-dimensional model experiment

## 3.2 Experimental procedure

Fig.14 illustrates the geometric parameters of the mold and boreholes. The model dimensions are 400 mm × 200 mm, and the mold is fabricated by cutting a corrugated pipe with a diameter of 400 mm. The cut zone is designed at the central position on the top surface of the model, adopting a square borehole layout. The geometric center of the square aligns with the geometric center of the top surface of the model, and the square's side length is 80 mm. The cut hole is located at the center of the square, with a diameter of 6 mm. The depths of the cut holes for models M-1 to M-5 are 50 mm, 55 mm, 60 mm, 65 mm, and 70 mm, respectively. The non-cut holes are positioned at the four vertices of the square, each with a diameter of 6 mm and a depth of 50 mm.

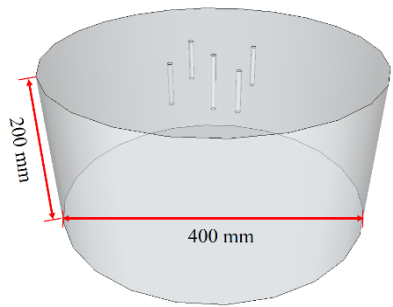

(a) Mold geometry parameters

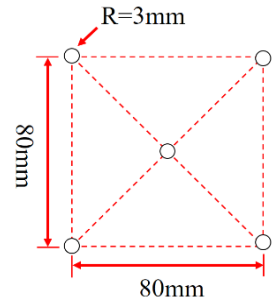

(b) Boreholes geometric parameters

Fig.14 Geometry parameters of mold and boreholes

Cement mortar was selected as the material for the model. The materials were weighed in accordance with a mix ratio of cement: sand: water = 1:7.25:1.55, thoroughly mixed using a mixer, and poured into molds for casting. A vibrator was used to ensure compaction and to remove air bubbles. The samples were cured under appropriate temperature and humidity conditions for 21 days. Simultaneously, a portion of the same mixture was poured into pre-prepared standard specimen molds measuring 50 mm × 100 mm. These specimens were also compacted using a vibration table to remove air bubbles and cured under the same conditions for 21 days. The cured standard specimens and models are shown in Fig.15. The basic physical and mechanical properties of the cured cement mortar standard specimens were determined using a CMT5000 universal testing machine and a DPO 5104B digital oscilloscope. The measured and calculated results are provided in Table 1.

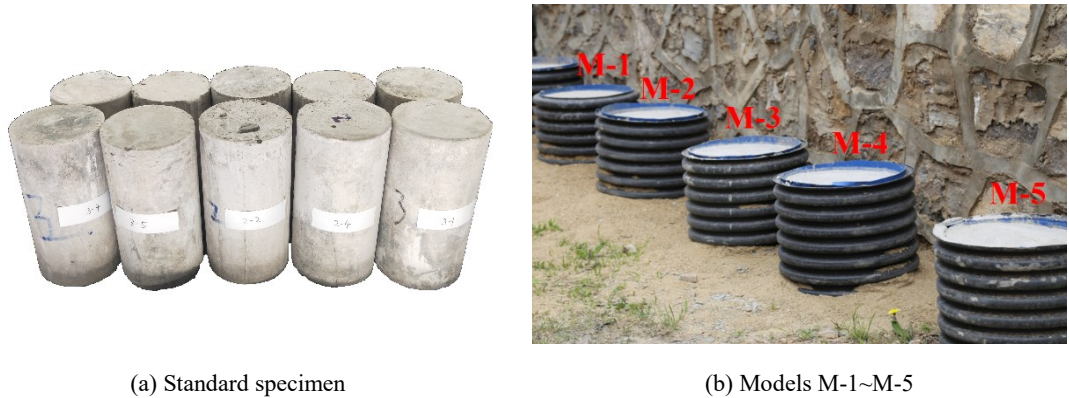

(a) Standard specimen　　　　　　　　　　　(b) Models M-1~M-5

Fig.15 Standard specimen and models M-1~M-5

Table1 Basic physical and mechanical properties of cement mortar

| $\rho/(g\cdot cm^{-3})$ | $\sigma_c/(MPa)$ | $E/(GPa)$ | $\mu$ | $C_p/(m\cdot s^{-1})$ |
|---|---|---|---|---|
| 1.97 | 14.5 | 8.01 | 0.28 | 3279 |

The physical and schematic diagrams of the explosive are shown in Fig.16. The borehole depth is 50 mm, with a charge factor of 0.7, a charge length of 35 mm, and a cartridge diameter of 6 mm. The explosive used is Lead Azide, and the end of initiation probe is positioned at the bottom of the explosive for detonation. The process of preparing the explosive cartridge involves the following steps: first, the initiation probe is fixed at the bottom of a plastic tube using hot glue; after the glue fully cools, 1.70 g of explosive is loaded, and finally, plasticine is used to seal the borehole.

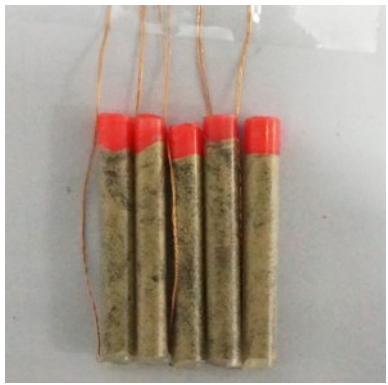 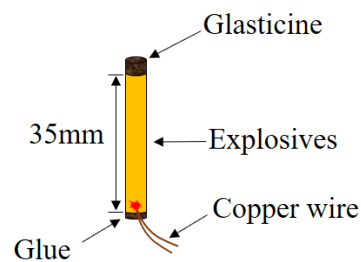

(a) Image of the explosive　　　　　　　　　(b) Schematic diagram of the explosive

Fig.16 Explosive

Fig.17(a) shows the prefabricated borehole mold. The mold was prepared by fixing five iron rods with a diameter of 6 mm onto a 500 mm × 200 mm wooden board, based on the borehole geometric parameters shown in Fig.14(b). A separate mold was made for each model, with the central iron rod in the molds having lengths of 50 mm, 55 mm, 60 mm, 65 mm, and 70 mm, respectively. The borehole mold was placed at the center of the model before the mortar solidified. After the mortar initially set for 2~3 hours, the mold was removed, and the boreholes were sealed with plastic bags to reduce the likelihood of foreign objects entering the boreholes. The processed boreholes are shown in Fig.17(b). Before the experiment, a hand drill was used to clear debris from the boreholes (Fig.17(c)), followed by charging and wiring the explosives. The model with completed charging and wiring is shown in Fig.17(d).

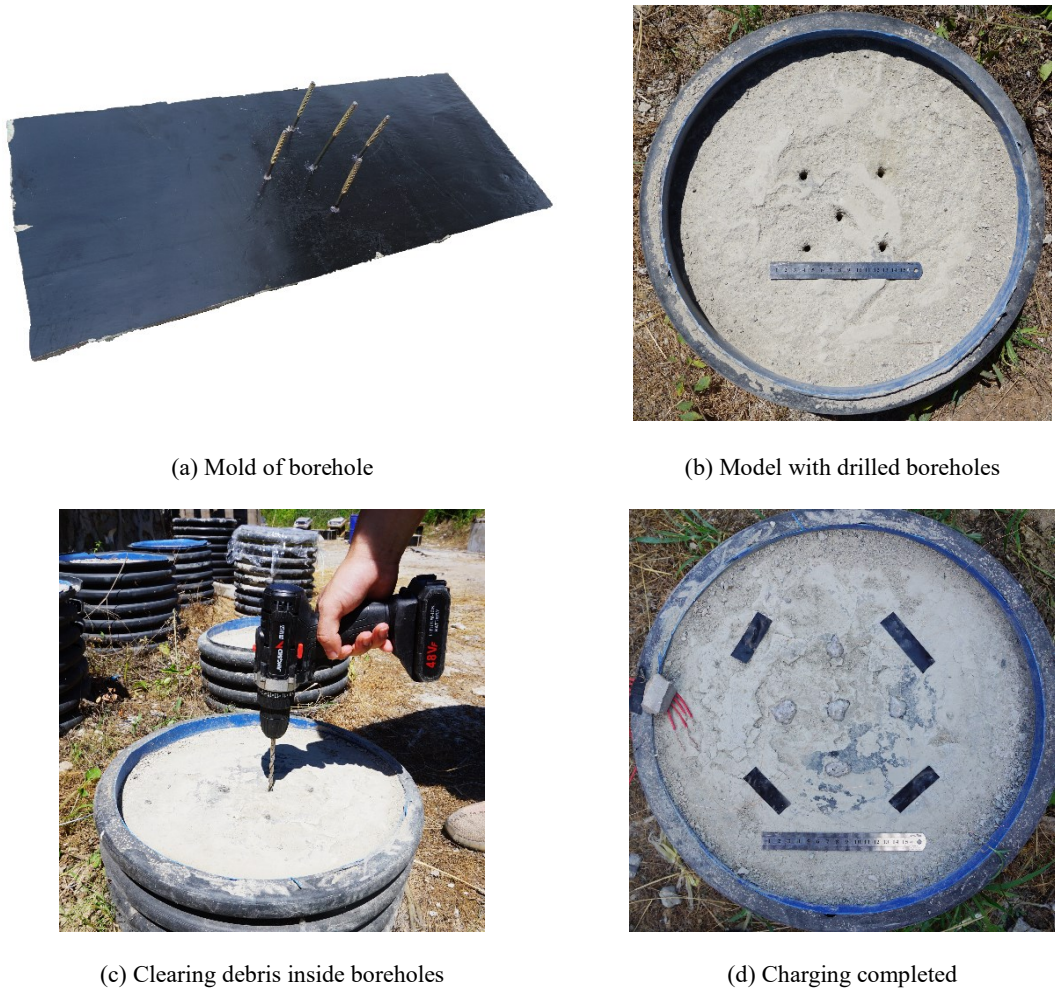

(a) Mold of borehole  (b) Model with drilled boreholes

(c) Clearing debris inside boreholes  (d) Charging completed

Fig.17 Drilling and charging

## 3.3 Analysis of post-blast result
### 3.3.1 Analysis of borehole utilization efficiency

Fig.18 illustrates the measurement process and calculation results of borehole utilization. After the blasting experiment, the setup was ventilated for 10 minutes to allow the removal of blasting fumes. Subsequently, the surface of the model and the debris within the cavity were cleared to expose the bottom of the boreholes. The distance from the bottom of each borehole to the free surface was measured individually, and the average value was taken as the cavity depth for the experiment. The ratio of the cavity depth to the auxiliary hole depth was calculated to determine the borehole utilization for this experiment. The borehole utilization results for model experiments M-1 to M-5 are presented in Fig.18(b).

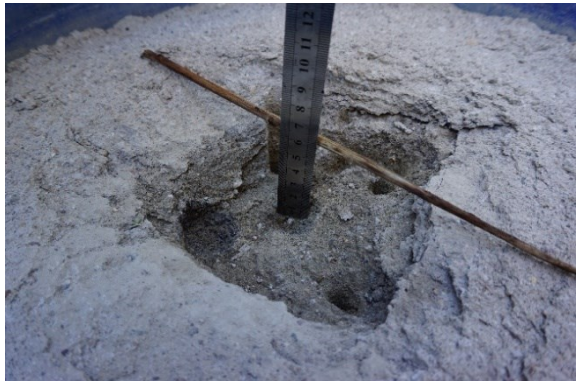 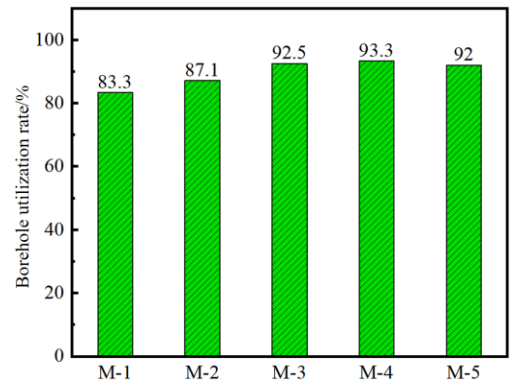

(a) Image of measuring boreholes depth    (b) Calculation results of borehole utilization rate

Fig.18 Measurement process and calculation results of borehole utilization rate

The borehole utilization ranks in descending order as M-4 > M-3 > M-5 > M-2 > M-1. As the extra-depth increases, borehole utilization initially rises to a peak and then declines. The cut blasting performance is optimal for M-3, with an extra-depth of 10 mm, and worst for M-1, with an extra-depth of 0 mm. The reasons can be analyzed as follows: when the extra-depth is small, the minimum resistance line of the cut hole does not reach the critical minimum resistance line. Consequently, the explosive energy in the cut hole is not fully utilized, resulting in shallow cavity depth and small cavity volume. This inadequacy leads to insufficient free surfaces and expansion space for the subsequent non-cut holes. As the extra-depth increases, the minimum resistance line of the cut hole approaches the critical minimum resistance line, enhancing cavity depth and volume and providing sufficient free surfaces and expansion space for the non-cut holes. However, when the extra-depth further increases beyond the critical minimum resistance line, the confinement effect at the bottom of the borehole leads to the emergence of residual borehole phenomena. This reduces the overall borehole utilization. If the extra-depth were infinitely large, the blasting of the cut hole would no longer influence the non-cut holes, and the latter would lose the free surfaces and expansion space provided by the cut hole. This implies that with an infinitely large extra-depth, the overall borehole utilization would be minimal, explaining the decline observed in borehole utilization at larger extra-depths.

### 3.3.2 Analysis of cavity diameter

Fig.19 illustrates the measurement process and results for the cavity diameter. After the experiment, the maximum and minimum diameters of the cavity were measured using a ruler, and their average value was calculated to determine the cavity diameter. As shown in the figure, the cavity diameter exhibits a trend of initially increasing and then decreasing with the increase in extra-depth. Under the experimental condition of M-4, the cavity diameter reaches its maximum value of 141 mm, while under the M-1 condition, it is at its minimum value of 122 mm.

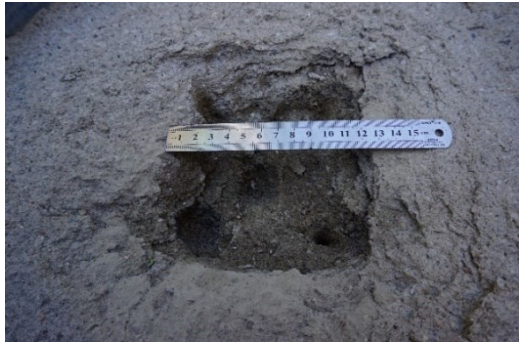
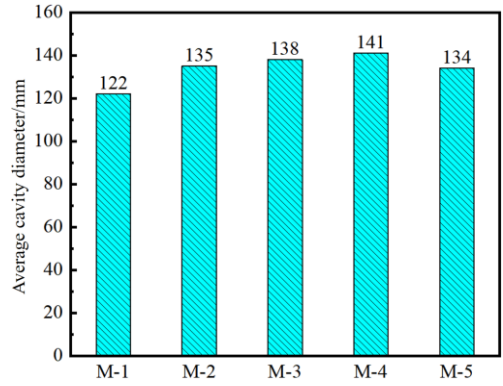

(a) Image of measuring cavity diameter       (b) Measuring results of cavity diameter

Fig.19 Measurement process and results of cavity diameter

### 3.3.3 Analysis of cavity volume

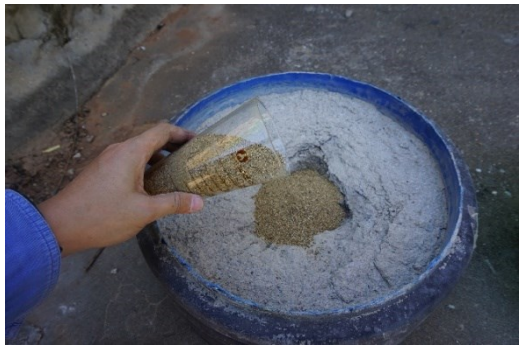
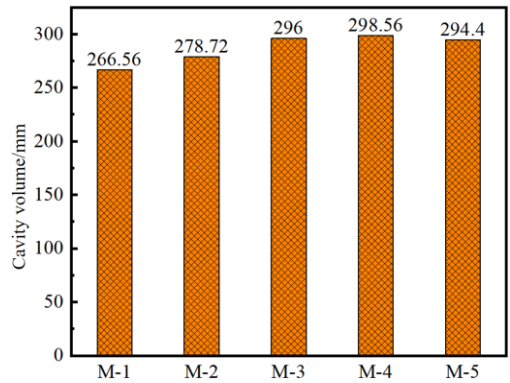

(a) Image of measuring cavity volume        (b) Measuring results of cavity volume

Fig.20 Measurement process and results of cavity volume

Fig.20 illustrates the measurement process and results of the cavity volume. After completely removing all debris from the cavity, fine sand was poured into the cavity using a graduated cylinder. The volume of the fine sand approximates the cavity volume. As the extra-depth increases, the cavity volume exhibits a trend of gradual growth, reaching a peak, and then decreasing. This trend aligns with the variations observed in the cavity diameter and borehole utilization rate. The underlying reasons for this phenomenon have been explained in detail in the subsection on borehole utilization analysis and will not be reiterated here.

### 3.3.4 Analysis of fragment size

The screening equipment used and an example of the screening results are shown in Fig.21. After the experiment, the blasted rock fragments were collected and subjected to a sieving test using gravel sieves that comply with national standards, which specify sieve opening sizes of 9.50, 16.00, 19.00, 26.50, 31.50, 37.50, and 53.00 mm. The mass of rock fragments in different gradation categories was measured and recorded after sieving.

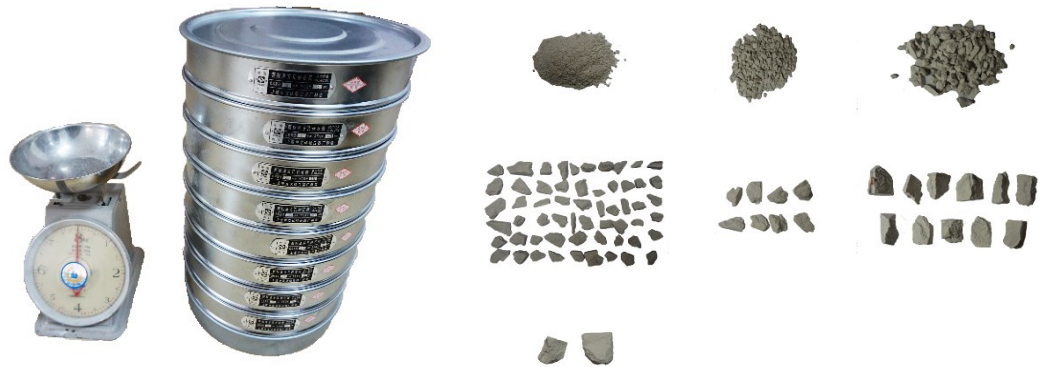

(a) Sieving equipment　　　　　　　　(b) Screening sample

Fig.21 Screening equipment and screening samples

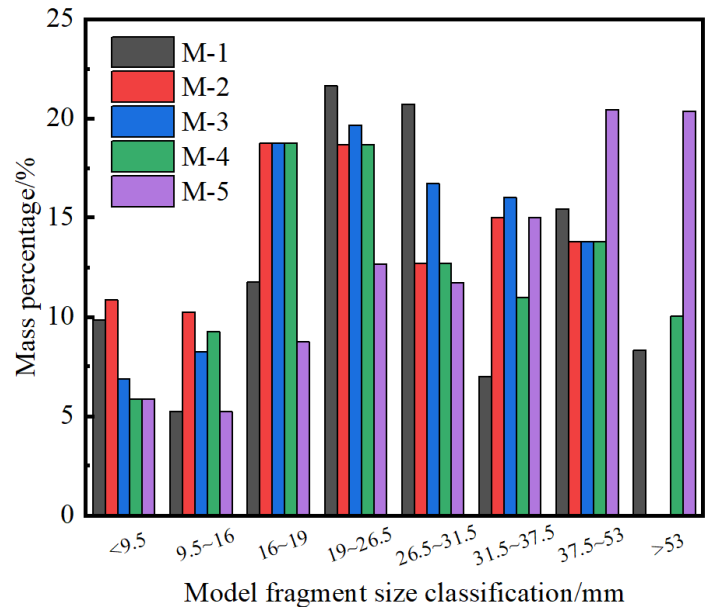

Fig.22 Fragment block size distribution histograms for M-1~M-5

　　Fig.14 shows the histograms of blast fragmentation distribution for M-1 to M-5 based on the sieving results. It can be observed from the figure that for M-1, where the cut hole depth is 50 mm, the mass of rock fragments with a particle size exceeding 37.5 mm accounts for 30.8% of the total rock mass. For M-2, where the cut hole depth is 55 mm, this proportion is 28.8%. For M-3, where the cut hole depth is 60 mm, the proportion is 29.8%. For M-4, where the cut hole depth is 65 mm, the proportion increases to 34.5%. For M-5, where the cut hole depth is 70 mm, the proportion significantly increases to 55.8%. Analyzing from the perspective of the coarse fragment ratio, it is evident that the coarse fragment ratio increases with the extra-depth. The reasons for this trend are as follows: when the extra-depth is small, the minimum resistance line of the cut hole does not reach the critical minimum resistance line, and more explosive energy in the cut hole is used for ejecting rocks rather than being effectively utilized, resulting in a smaller cavity volume and a higher proportion of small rock fragments. As the extra-depth continues to increase and the minimum resistance line approaches the critical minimum resistance line, the explosive energy in the cut hole is reasonably utilized, with a portion used for rock breakage and another portion for rock ejection, leading to a more uniform fragmentation size. However, when the cut hole depth continues to increase, the minimum resistance line exceeds the critical minimum resistance line. Due to the clamping effect at the hole bottom, residual boreholes begin to appear. The explosive energy in the cut hole becomes insufficient to eject the rock, forming larger through-going fractures in the rock mass and increasing the coarse fragment ratio.

## 4 Discussion

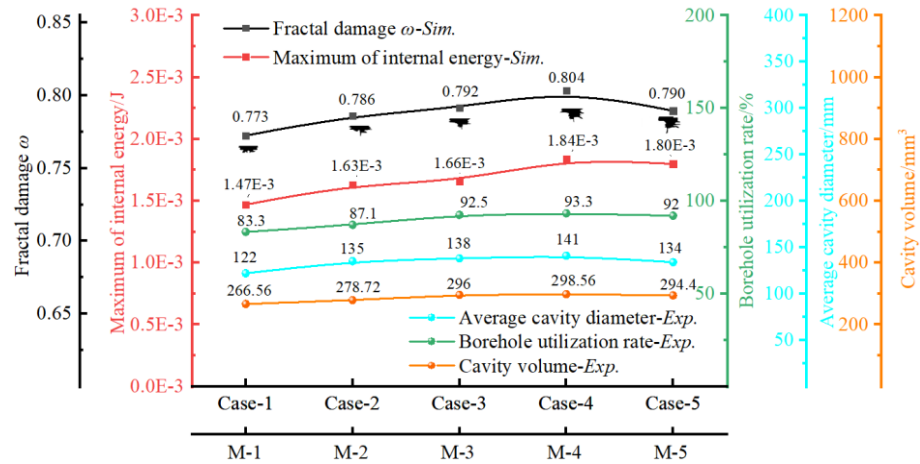

Fig23. Summary of the curves between various blasting damage parameters and extra-depth in simulations and experiments

Fig.23 summarizes the relationships between various blasting damage parameters and extra-depth as observed in simulations and experiments. The figure shows that as the extra-depth increases, the fractal damage and the maximum internal energy obtained from 2D numerical simulations exhibit an overall trend of rising initially, followed by a slight decline. Similarly, the results from the 3D model experiments reveal analogous patterns in post-blasting parameters such as borehole utilization, cavity diameter, and cavity volume. These findings, derived from both numerical simulations and physical experiments, illustrate the nonlinear influence of extra-depth on blasting damage. As the extra-depth of the cut hole increases, all blasting damage parameters demonstrate a consistent pattern of initial growth followed by a decline. This study also provides corroborative evidence that fractal damage characterization and 2D numerical simulation methods are applicable and effective for predicting post-blasting damage in 3D model experiments.

## 5 Conclusions

This study systematically investigates the rock fragmentation mechanism of extra-depth cut blasting using a combination of LS-DYNA 2D numerical simulations and 3D cement mortar model experiments. Based on fractal damage theory and various experimental measurement methods, the research provides a scientific theoretical foundation for optimizing extra-depth parameters in deep-hole blasting. The main conclusions are as follows:

(1) Nonlinear impact of extra-depth on blasting damage validated through 2D simulations and 3D experiments: Both 2D numerical simulations and 3D model experiments consistently demonstrate the nonlinear effects of extra-depth on blasting damage. In 2D simulations, as the extra-depth of the cut hole increases, fractal damage and maximum internal energy exhibit a trend of initial increase followed by a decrease. Similarly, in 3D experiments, parameters such as borehole utilization, cavity diameter, and cavity volume follow the same pattern. These findings indicate that a moderate increase in extra-depth enhances rock fragmentation efficiency, but excessive extra-depth leads to decreased blasting efficiency due to confinement at the hole bottom.

(2) Optimal extra-depth and energy allocation between fragmentation and throw: The optimal extra-depth significantly influences the allocation of explosive energy between fragmentation and throw. In 2D simulations, fractal damage and internal energy reach their peak when the extra-depth is 15 mm. In 3D experiments, the cavity size and borehole utilization reach maximum values when the extra-depth is 10 mm. Although the exact critical values differ, both approaches confirm the existence of a minimum critical extra-depth. When the cut hole depth aligns with this critical extra-depth, explosive energy achieves an optimal balance between fragmentation and throw, maximizing cut blasting efficiency.

(3) Fractal damage as a proxy for internal energy and quantitative damage evaluation: Fractal damage and internal energy show a high degree of correlation, offering a new perspective for quantitative post-blast damage evaluation. The numerical simulation results reveal that during the later stages of blasting, the maximum internal energy within the model correlates strongly with the trend of fractal damage as extra-depth varies. Similarly, 3D experiments demonstrate that the quantified parameters—borehole utilization, cavity diameter, and cavity volume—correspond to the fractal damage trends. These findings suggest that beyond directly calculating fractal damage, internal energy in numerical simulations can serve as a quantitative indicator for evaluating post-blast damage. This approach provides a novel pathway for rapid damage assessment in engineering practice.

## Acknowledgements

Changda Zheng: Experimental device design, Software, Writing– original draft, Writing– review & editing. Renshu Yang: Investigation, Resources. Jinjing Zuo and Canshu Yang: Software support and Visualization. Yuanyuan You: Laboratory support and Experiment support. Zhidong Guo: Investigation.

## Declaration of competing interest

The authors declare that they have no known competing financial interests or personal relationships that could have appeared to influence the work reported in this paper.

## Conflict of Interest

The authors would like to acknowledge the anonymous reviewers for their valuable and constructive comments. This work was financially supported by the National Natural Science Foundation of China (No. 52204085 and No. 51934001).